\documentclass[aps,twocolumn,showpacs,prl,twoside,amssymb,amsmath]{revtex4}
\usepackage{amssymb}
\usepackage{graphicx}
\usepackage{amsmath}
\usepackage{colordvi}
\usepackage{bbm}
\usepackage{amsmath}
\newcommand{\tr}{{\rm Tr}}
\newcommand{\mC}{{\mathcal C}}
\begin{document}

\title{Tight bounds for the quantum discord}

\author{Sixia Yu$^{1,2}$}
\email{cqtys@nus.edu.sg}
\author{Chengjie Zhang$^{1}$}
\author{Qing Chen$^{1,2}$}
\author{C.H. Oh$^{1}$}
\email{phyoh@nus.edu.sg}
\affiliation{$^1$Centre for quantum technologies and Physics department, National University of Singapore, 3 Science Drive 2, Singapore 117543\\
$^2$Hefei National Laboratory for Physical Sciences at Microscale and Department of Modern Physics, \\
University of Science and Technology of China, Hefei, Anhui 230026, China}
\begin{abstract}
Quantum discord quantifies quantum correlations beyond entanglement and assumes nonzero values, which are notoriously hard to compute, for almost all quantum states. Here we provide computable tight bounds for the quantum discord for qubit-qudit states. In the case of two qubits our lower and upper bounds coincide for a 7-parameter family of filtered $X$-states, whose quantum discords can therefore be evaluated analytically. An application to the accessible information of the binary qubit channel is also presented. For the qubit-qudit state output by the  circuit of deterministic computation with one qubit, nontrivial lower and upper bounds that respect the zero-discord conditions are obtained for its quantum discord.

\end{abstract}


\maketitle
The quantum discord \cite{discord1,discord2} has gradually become another important resource for quantum informational processing tasks besides the entanglement. Firstly, in certain quantum mechanical tasks such as the deterministic quantum computation with one qubit (DQC1) \cite{DQC1}, the quantum advantages can be gained  \cite{Datta} with the presence of quantum discord while the entanglement is absent. Secondly the quantum discord is shown to be a better indicator of the quantum phase transition in certain physical systems than the entanglement \cite{pt}.
Thirdly, in addition to its original interpretation via Maxwell demon  \cite{demon}, the operational interpretations of quantum discord via state merging \cite{merg} establish firmly the status of the quantum discord as an essential resource amidst other concepts of quantum information.

As a measure of the quantum correlation beyond entanglement, the quantum  discord of a given state $\varrho$ of a composite system $AB$ is \cite{discord1,discord2}
\begin{equation}\label{def}
D_A(\varrho):=\min_{\{E_i^A\}}\sum_{i}p_iS(\varrho_{B|i})+S(\varrho_A)-S(\varrho),
\end{equation}
where $S(\varrho)=-\mathrm{Tr}(\varrho\log_2\varrho)$ denotes the von Neumann entropy and the minimum is taken over all the positive operator valued measures (POVMs) $\{E_i^A\}$ on the subsystem $A$ with $p_i=\tr (E_i^A\varrho)$ being the probability of the $i$-th outcome and $\varrho_{B|i}=\tr_A(E_i^A\varrho)/p_i$ being the conditional state of subsystem $B$. The minimum can also be taken over all the von Neumann measurements~\cite{discord1}. These two definitions are in general inequivalent even for qubits and our proposed bounds in Eq.(\ref{bds}) below apply to both of them. When the measurements are made on subsystem $B$ the quantum discord $D_B(\varrho)$ can be defined similarly and is in general different from $D_A(\varrho)$.

Quantum discord assumes nonnegative values and zero-discord states are relatively well understood: $D_A(\varrho)=0$ if and only if there exists a complete orthonormal basis $\{|k\rangle\}$ for the subsystem $A$ together with some density matrices $\varrho_k^B$ for the subsystem $B$ such that $\varrho=\sum_k p_k |k\rangle\langle k|\otimes\varrho_k^B$. Recently various methods to detect zero discord \cite{cond1,condition} have been proposed for a given state as well as for an unknown state \cite{witness}. Besides its initial motivation in pointer states in measurement problem \cite{discord1}, vanishing quantum discord is also found to be related to the complete positivity of a map \cite{CP} and the local broadcasting of quantum correlations \cite{broadcast}.

Unfortunately zero-discord states are of zero measure \cite{almost} and the nonzero values of the quantum discord are notoriously difficult to compute because of the minimization over all the possible measurements. There are only  a few analytical results including the Bell-diagonal states  \cite{bells} and rank-2 states \cite{r2}, in addition to a thorough numerical calculation \cite{max} in the case of von Neumann measurements.
For two-qubit $X$-states, since there are counter examples \cite{chen,lu} for the algorithm proposed in \cite{X2}, the evaluation of their quantum discords remains a nontrivial task. It is therefore desirable to have some computable bounds for the quantum discord so that the question of how large or small the quantum discord can possibly be, e.g., in the DQC1 circuit, can be answered more reasonably.

In this Letter we shall provide computable tight  bounds for the quantum discord of qubit-qudit states. For a family of two-qubit filtered $X$-states with 7 parameters up to local unitary transformations (LUTs) our lower and  upper bounds coincide so that their quantum discords can be evaluated analytically. Also we present an application to the accessible information of the binary qubit channel for which our lower and upper bounds can coincide. For the quantum discord of the qubit-qudit state output by the DQC1 circuit we derive nontrivial lower and upper bounds, which qualitatively agree with the zero discord conditions \cite{cond1,witness} comparing with the estimation in~\cite{Datta}.

To present our main result we need some notations. First of all, we denote by $\varrho$ a qubit-qudit state  with the reduced density matrices $\varrho_A$ and $\varrho_B$ for the qubit $A$ and the qudit $B$ respectively.  Let $\vec x$ be the Bloch vector for $\varrho_A$ with its length given by $x^2=2\tr\varrho^2_A-1$. Without loss of generality we assume that the reduced density matrix $\varrho_A$ of qubit $A$, on which the measurement is performed, is invertible since otherwise we have a product state, which has zero discord. Therefore the following filtered density matrix is well defined:
\begin{equation}\label{tilde}
\tilde\varrho=\frac1{\sqrt{2\varrho_A}}\ \varrho\ \frac1{\sqrt{2\varrho_A}}.
\end{equation}

Secondly, let $\sigma_\mu$ $(\mu=0,1,2,3)$ be the identity matrix and three standard Pauli matrices.  We associate with every qubit-qudit state $\varrho$ (or $\tilde\varrho$) a positive semi-definite two-qubit operator
\begin{equation}\label{Q}
2\tr_{B_1B_2}[(1-V^B_{12}) \varrho\otimes\varrho]=\frac14\sum_{\mu,\nu=0}^3[Q_\varrho]_{\mu\nu}\sigma_\mu\otimes\sigma_\nu
\end{equation}
in which $V^B_{12}=\sum_{ij}|ij\rangle\langle ji|$ stands for the two-qudit swapping operator. Then the equation $\det({Q}_{\varrho}-q\eta)=0$, where $\eta={\rm diag}(1,-1,-1,-1)$,  has four real solutions $q_1\ge q_2\ge q_3\ge q_4$ \cite{hil,ulm}.
It should be noted that each $q_i$ is as readily computable as the concurrence of $\varrho$, which equals to $\max\{0,\sqrt{q_2}+\sqrt{q_3}\pm \sqrt{q_4}-\sqrt{q_1}\}/2$ \cite{filter}, in the case of two-qubit states for which we have $Q_{\varrho}=R_\varrho\eta R_\varrho^T$ with $R_\varrho$ being the $4\times4$ correlation matrix defined by $[R_\varrho]_{\mu\nu}=\langle\sigma_\mu\otimes\sigma_\nu\rangle_\varrho$. These values $q_i(\varrho)$, showing explicitly the dependence on $\varrho$, are invariant under an arbitrary Lorentz transformation (LT) $L$, satisfying ${L}\eta {L}^T=\eta$, that brings $Q_{\varrho}$ to $LQ_{\varrho} L^T$. The local filter $\sqrt[4]{1-x^2}/\sqrt{2\varrho_A}$ acting on qubit $A$ brings $Q_\varrho$ to $(1-x^2)Q_{\tilde\varrho}$ and induces an LT to $Q_\varrho$, from which it follows that $(1-x^2)q_2(\tilde\varrho)=q_2(\varrho)$.

Thirdly, we denote by $Q_{\tilde\varrho}^{3\times3}$ the $3\times 3$ matrix obtained from the $4\times4$ matrix $-Q_{\tilde\varrho}$ by deleting its first row and column. Let $t_1(\tilde\varrho)$ be the largest eigenvalue of $Q_{\tilde\varrho}^{3\times3}$ corresponding to the eigenvector ${\vec e}$. We define  $\vec m_{\tilde\varrho}$ to be the unit  vector along the direction $\sqrt{1-x^2}\vec e_\perp+\vec e_\parallel$, where $\vec e_\perp=\vec e-\vec e_\parallel$ and $\vec e_\parallel=\vec x(\vec x\cdot\vec e)/x^2$. In the case of $x=0$ we define $\vec m_{\tilde\varrho}=\vec e$. By measuring the observable ${\vec m_{\tilde\varrho}\cdot\vec \sigma}$ on the qubit $A$ we obtain a suboptimal value $D_A(\varrho|{\vec m}_{\tilde\varrho})$ for the quantum discord as in Eq.(\ref{def}) without minimization. For a two-qubit state $\varrho$, since $\tr_B\tilde\varrho=\sigma_0/2$, we have $Q_{\tilde\varrho}^{3\times 3}={T}_{\tilde\varrho}{T}_{\tilde\varrho}^T$  where $T_{\tilde\varrho}$ is the $3\times 3$ correlation matrix for $\tilde\varrho$ defined by $[T_{\tilde\varrho}]_{ab}=\langle\sigma_a\otimes\sigma_b\rangle_{\tilde\varrho}$ $(a,b=1,2,3)$.

Lastly, we need to introduce an increasing convex function $co(1-z^2)$ of $z$ where
\begin{equation}
co(z)=\left\{\begin{array}{lc} h(z), &\quad \mbox{if }z\ge 0,\\
\log_2\frac2{1+z},&\quad \mbox{if }z\le 0,\end{array}\right.
\end{equation}
and $h(z)=-\frac{1+\sqrt{z}}2\log_2\frac{1+\sqrt{z} }2-\frac{1-\sqrt z}2\log_2\frac{1-\sqrt{z}}2$. For any density matrix $\varrho$ it holds $S(\varrho)\ge co(2\tr\varrho^2-1)$  since  we always have $S(\varrho)\ge -\log_2\tr\varrho^2$ and if $\tr\varrho^2\ge 1/2$ we have $S(\varrho)\ge h(2\tr\varrho^2-1)$ which can be read off from the information diagram between the entropy and index of coincidence \cite{ci}.

{\it Theorem  } For a qubit-qudit state $\varrho$, with subsystem $A$ being a qubit, the following bounds hold
\begin{equation}\label{bds}
co({\mathcal L})+S(\varrho_A)-S(\varrho)\le  D_A(\varrho)\le D_A(\varrho|{\vec m_{\tilde\varrho}}),
\end{equation}
where ${\mathcal L}=2\tr\varrho_B^2-1+q_2(\varrho)$. In the case of two qubits, the lower and upper bounds coincide  if $q_2(\tilde\varrho)=t_1(\tilde\varrho)$.

A detailed proof can be found in \cite{online} and an outline is given in what follows.
The upper bound holds true by definition. The key to prove the lower bound is the Koashi-Winter relation \cite{rel}:
$D_A(\varrho_{AB})+S(\varrho_{AB})-S(\varrho_{A})=E_F(\varrho_{BC})$, where $E_F(\varrho_{BC})$ is the
entanglement of formation \cite{eof} of the $d\times 2d$ state $\varrho_{BC}=\tr|\psi\rangle\langle\psi|_{ABC}$  with  $|\psi\rangle_{ABC}$ being a purification of  $\varrho_{AB}$ in a $2\times d\times 2d$ system $ABC$.
The lower bound follows from $E_F(\varrho_{BC})\ge co(1-\mC^2_{BC})$, where $\mC_{BC}$ is the concurrence of $\varrho_{BC}$, and an evaluation of the concurrence $\mC_{BC}^2=1-{\mathcal L}$ by applying {\it Theorem 4} in \cite{ulm}. Furthermore  the ensemble of $\varrho_{BC}$ induced by measuring $\vec m_{\tilde\varrho}\cdot\vec \sigma$ on qubit $A$ is optimal for the tangle $\tau_{BC}$ of $\varrho_{BC}$ whose value  $2(1-\tr\varrho_B^2)-(1-x^2)t_1(\tilde\varrho)$ can be readily obtained by applying {\it Theorem 1} in \cite{osb}. Since $h(1-z)$ is a concave function of $z$ we have bound $D_A(\varrho|\vec m_{\tilde\varrho})\le h(1-\tau_{BC})+S(\varrho_A)-S(\varrho_{AB})$, which coincides with the lower bound if $q_2(\tilde\varrho)=t_1(\tilde\varrho)$, i.e., $\tau_{BC}=\mC_{BC}^2$, since ${\mathcal L}\ge0$ in the case of two qubits.

\begin{figure}
\includegraphics[scale=0.4]{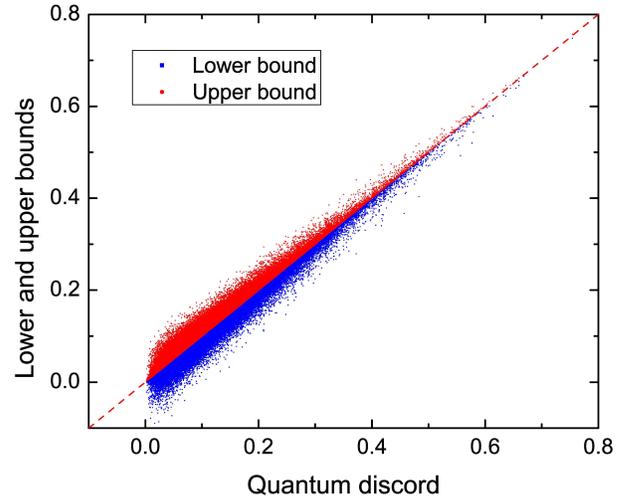}
\caption{(Color online) Comparing the upper and lower bounds with the quantum discord minimized over von Neumann measurements for $10^{5}$ randomly chosen 2-qubit states of rank 4.}
\end{figure}
For two-qubit states we have carried out some numerical calculations and a comparison with our bounds is shown in Fig.1, where for convenience the upper bound is taken to be $h(1-\tau_{BC})+S(\varrho_A)-S(\varrho_{AB})$, which is slightly weaker than Eq.(\ref{bds}). For about 70\% of  $10^5$ randomly chosen states of rank 4 the differences between our bounds and the values of quantum discord obtained by minimization over von Neumann measurements lie in the range of  $\pm0.01$ with a maximal difference about $\pm0.1$. We note that for some states our lower bounds are negative and therefore are trivial. On the other hand there exist several nontrivial families of states, for which the upper and lower bounds coincide so that analytical expressions of quantum discord for these states can be obtained. In these cases the two definitions of quantum discord become identical.

{\it Example 1: Bell-diagonal states \cite{bells}.} A Bell-diagonal state has a density matrix  $\varrho_{BS}=\frac14\sum_\mu c_\mu\sigma_\mu\otimes\sigma_\mu$ with  $c_\mu$ being real and $|c_\mu|\le c_0=1$. It is clear that $\tilde\varrho_{BS}=\varrho_{BS}$ and ${Q}_{\varrho_{BS}}={\rm diag}(1,-c_1^2,-c_2^2,-c_3^2)$. As a result we have $q_2(\tilde\varrho_{BS})=t_1(\tilde\varrho_{BS})=\max\{c_1^2,c_2^2,c_3^2\}$ so that upper and lower bounds as in Eq.(\ref{bds}) coincide.

{\it Example 2: Rank-2 states \cite{r2}. } Every two-qubit state of rank 2 admits a 3-qubit purification $|\psi\rangle_{ABC}$ with the reduced density matrix $\varrho_{BC}$ being a two-qubit state of rank 2. As a result $\tau_{BC}=\mC^2_{BC}$ so that the upper and lower bounds as in Eq.(\ref{bds}) coincide.

{\it Example 3: $X$-states.} In many scenarios $X$-state arises as the two-particle reduced density matrix as long as there is a certain symmetry of the physical system. In general a 2-qubit $X$-state
\begin{equation}
{\varrho_X}=\left(\begin{array}{cccc} \varrho_{00}&0&0&\varrho_{03}\cr 0&\varrho_{11}&\varrho_{12}&0\cr
0&\varrho_{21}&\varrho_{22}&0\cr \varrho_{30}&0&0&\varrho_{33}\end{array}\right)
\end{equation}
has 7 independent real parameters. Since the quantum discord is invariant under LUTs, we can assume without loss of generality that $\varrho_{03}$ and $\varrho_{12}$ are real and therefore we have in fact only 5 real parameters, which can be conveniently taken as those nonzero entries of the correlation matrix $R_{\varrho_X}$, i.e., $x=\langle\sigma_3\otimes\sigma_0\rangle_{\varrho_X}$, $y=\langle\sigma_0\otimes\sigma_3\rangle_{\varrho_X}$, and $s_a=\langle\sigma_a\otimes\sigma_a\rangle_{\varrho_X}$ ($a=1,2,3$). Since  $Q_\varrho=R_\varrho\eta R_\varrho^T$  holds for qubits it is easy to see that four solutions to the equation $\det(Q_{\varrho_X}-q\eta)=0$ are $ s_{1,2}^2$ and $4(\sqrt{\varrho_{00}\varrho_{33}}\pm\sqrt{\varrho_{11}\varrho_{22}})^2$. Furthermore the $3\times3$ correlation matrix ${T}_{\tilde\varrho_X}$ is diagonal with entries $s_{1,2}/\sqrt{1-x^2}$ and $(s_3-xy)/(1-x^2)$. Thus
as long as
\begin{equation}\label{cond}
|\sqrt{\varrho_{00}\varrho_{33}}-\sqrt{\varrho_{11}\varrho_{22}}|\le |\varrho_{03}|+|\varrho_{12}|
\end{equation}
we have $q_2(\tilde\varrho_X)=t_1(\tilde\varrho_X)=\max \{s^2_{1,2}\}/(1-x^2)$ so that
the upper and lower bounds of the quantum discord coincide. We denote by $\varrho_{X_c}$ the $X$-state that satisfies the condition Eq.(\ref{cond}) and the optimal observable to measure is $\sigma_1$ if $s_1\ge s_2$ and $\sigma_2$ otherwise. Explicitly,
\begin{eqnarray}
 D_A(\varrho_{X_c})=h\left(y^2+\max \{s^2_{1,2}\}\right)+h(x^2)-\quad\quad\cr
 \quad\sum_{\pm} \frac{1\pm s_3}2h\left(\frac{(x\pm y)^2+(s_1\mp s_2)^2}{(1\pm s_3)^2}\right).
\end{eqnarray}

{\it Example 4: Filtered $X$-states.}  Take an arbitrary 2-qubit state $\varrho$ satisfying i) $q_2(\tilde\varrho)=t_1(\tilde\varrho)$, i.e., the upper and lower bounds for $D_A(\varrho)$ coincide, and ii) $\varrho_A$ is proportional to identity, for example the state $\varrho_{X_c}$ with $x=0$ as in example 3, and take an arbitrary invertible Hermitian operator $F_A$ acting only on qubit $A$. Then for the filtered state $\varrho_F\propto {F_A\varrho F_A^\dagger}$
we have $\tilde\varrho=\widetilde{\varrho_F}$ from the definition Eq.(\ref{tilde}) so that
$\varrho_F$ has also a coincided lower and upper bound. Because $cF_A$ and $F_A$ represent the same filter for any nonzero real $c$, there are $3$ independent real parameters for $F_A$. Thus we have a family of filtered $X$-states with 7 parameters for which the quantum discords can be evaluated analytically.

{\it Example 5: Accessible information.} A quantum communication channel $\{p_k,\varrho_k\}$ is defined by the action of sending states $\{\varrho_k\}$ with probabilities $\{p_k\}$ to a receiver who can perform any possible POVM to gain the information of $\{p_k\}$. The accessible information $I_{acc}$ is the maximum of the mutual information of the joint probability distribution $\{p_{ak}=p_k\tr E_a\varrho_k\}$ over all POVMs $\{E_a\}$. If we introduce a bipartite state $\varrho_{AB}=\sum_k p_k\varrho_k\otimes|k\rangle\langle k|$ where $\{|k\rangle\}$ is a complete orthonormal basis then we have $D_B(\varrho_{AB})=0$ and $D_A(\varrho_{AB})=\chi(\varrho_A)-I_{acc}$ where $\chi(\varrho_A):=S(\varrho_A)-\sum p_kS(\varrho_k)$ is the Holevo bound. For qubit channels the theorem provides an upper bound
$I_{acc}\le S(\varrho_B)-co({\mathcal L})$ as well as a lower bound given by the measurement along $\vec m_{\varrho}$ for the accessible information.

Consider a binary qubit channel $\{p_k,\varrho_k\}_{k=1}^2$ and denote by $\vec a$ and $\vec b$ the Bloch vectors of qubit states $\varrho_1$ and $\varrho_2$, respectively, and  $\vec c_\pm=p_1\vec a\pm p_2\vec b$. The lower bound for $I_{acc}$ provided by $\vec m_{\varrho}$ coincides with the lower bound ($M(t)$) in \cite{fuchs}. The $4\times4$ correlation matrix of the two-qubit state $\varrho=\sum_{k=1}^2 p_k\varrho_k\otimes|k\rangle\langle k|$  reads
\begin{equation}
R_{\varrho}=\left(\begin{array}{cccc}1&0&0&\delta\cr
\vec c_+&\vec0&\vec0&\vec c_-\cr\end{array}\right),
\end{equation}
where $\delta=p_1-p_2$. We note that $Q_{\varrho}=R_{\varrho}\eta R_{\varrho}^T$ is of rank 2 and therefore four solutions to the equation $\det(Q_{\varrho}-q\eta)=0$ are $\{0,0,(1-\delta^2)\lambda_\pm\}$, where
\begin{equation}
\lambda_\pm=\frac{1-\vec a\cdot\vec b\pm\sqrt{(1-a^2)(1-b^2)}}2,
\end{equation}
from which we obtain $q_2(\varrho)=(1-\delta^2)\lambda_-$. The $3\times3$ correlation matrix $T_{\tilde\varrho}$ of $\tilde\varrho$ is of rank 1 and the largest eigenvalue of $T_{\tilde\varrho}T_{\tilde\varrho}^T$ is $t_1(\tilde\varrho)=(1-\delta^2)^2\lambda_+\lambda_-/(1-c_+^2)^2$.
By solving $q_2(\tilde\varrho)=t_1(\tilde\varrho)$ we obtain $p_1^2(1-a^2)=p_2^2(1-b^2)$ in which case the upper and lower bounds for $I_{acc}$ coincide, meaning that the optimal POVM  for $I_{acc}$ is a von Neumann measurement (along the direction $\vec m_{\varrho}\propto\vec c_-$). This proves explicitly, for the first time as far as we know, that the optimal von Neumann measurement given in \cite{fuchs} is also optimal among POVMs.

 {\it Example 6: DQC1 states. }Let us consider the qubit-qudit state output by the DQC1 circuit, aiming at calculating efficiently the trace of a unitary operation $U$ acting on the qudit $B$, with a density matrix \cite{DQC1}
\begin{equation}
\varrho_{u}=\frac1{2d}\left(\begin{array}{cc}I& \alpha U^\dagger\\ \alpha U&I\end{array}\right)
\end{equation}
where $d=2^n$ and $0\le\alpha\le1$. Let $\{e^{i\phi_a}, |a\rangle\}_{a=1}^d$ be the eigensystem of $U$ and obviously $\varrho_u=\frac1d\sum_a\hat\phi_a\otimes |a\rangle\langle a|$ is separable where $\hat\phi_a=(1+\alpha \vec n_a\cdot\vec\sigma)/2$ with $\vec n_a=(\cos\phi_a,\sin\phi_a,0)$. The nonvanishing quantum discord of $\varrho_u$, which has been estimated in  \cite{Datta}, is argued to be responsible for the quantum speedup. Recently it is found that \cite{witness,cond1} $D_A(\varrho_u)=0$ if and only if $\alpha=0$ or $\beta=1$ where $\beta=(d+|\tr U^2|)/(2d)$. The estimation in \cite{Datta}, though respects the trivial condition $\alpha=0$, is insensitive to the condition $\beta=1$, i.e., a nonzero value of the quantum discord is estimated in this case, in which $U^2\propto I$ and $U$ can be in the same time typical, i.e., $u_1:=|\tr U|/d\approx0$. Our Theorem provides nontrivial bounds that respect both two zero-discord conditions.

For simplicity we shall assume $u_1=0$ first, i.e., $\sum\vec n_a=0$. In this case we  have $\tilde\varrho_u=\varrho_u$ and $S(\tr_B\varrho_u)=1$. It turns out that $Q_{\varrho_u}={\rm diag} \{2(1-1/d), -Q_{\varrho_u}^{3\times3}\}$ where $Q_{\varrho_u}^{3\times3}=\frac{2\alpha^2}{d^2}\sum_{a=1}^d\vec n_a\vec n_a$ has eigenvalues $\frac{2\alpha^2}{d}\{0,\beta,1-\beta\}$. As a result  we have $q_2(\varrho_u)=2\alpha^2\beta/d$ ($\beta\ge1/2$) which is also the largest eigenvalue of $Q_{\varrho_u}^{3\times3}$ with eigenvector $\vec m=(\cos\phi,\sin\phi,0)$ where  $e^{i2\phi}=\tr U^2/|\tr U^2|$. From Eq.(\ref{bds}) and ${\mathcal L}=2(1+\alpha^2\beta)/d-1\le 0$ for $d\ge 4$ it follows
\begin{equation}\label{b3}
\log_2\frac{2}{1+\alpha^2\beta}-h(\alpha^2)\le D_A(\varrho_u)\le h\left(\alpha^2\beta\right)-h(\alpha^2).
\end{equation}
In deriving the upper bound we have used $D_A(\varrho_u|\vec m)+h(\alpha^2)\le\frac1d\sum_a h(\alpha^2(\vec m\cdot \vec n_a)^2)\le h(\frac{\alpha^2}d\sum_a(\vec m\cdot\vec n_a)^2)$ in which the first inequality is valid for arbitrary $u_1$ and becomes an equality in the case of $u_1=0$. Thus the upper bound for $D_A(\varrho_u)$ given in Eq.(\ref{b3}) holds for all unitary $U$ and reaches its maximum $h(1/2)\approx 0.6$ at $\alpha=1$ and $\beta=1/2$.
It is obvious that in the case of $\alpha=0$ or in the case of pure qubit $\alpha=1$ and $\beta=1$ the lower and upper bounds coincide so that the quantum discord vanishes.

To conclude, in view of the hardship of computing the nonzero values of quantum discord, the computable tight bounds provided here for qubit-qudit states should be useful in further quantitative studies of the relation between quantum discord and phase transitions, quantum speedups, and so on. Notably our bounds enable us to evaluate analytically the quantum discords of a family of filtered $X$-states with 7 parameters up to LUTs (for comparison there are 9 parameters for a general 2-qubit state) and to estimate the quantum discord in the DQC1 circuit more reasonably. Also our bounds are applicable for the classical correlation and accessible information. Though we have restricted  to qubit-qudit systems, for which the concurrence $\mC_{BC}$ can be evaluated exactly, the bounds for the quantum discord $D_A(\varrho_{AB})$ of  a general bipartite state, which might not be specially tight, can be obtained in a similar way via the bounds for the entanglement of formation $E_F(\varrho_{BC})$.

We acknowledge the financial supports of A*STAR Grant No. R-144-000-189-305, CQT project WBS: R-710-000-008-271, and NNSF of China (Grant No. 11075227).


\section{Supplementary material: Proof of the theorem}

Let $\{|\psi_i\rangle\}_{i=1}^{2d}$ be the eigenstates of a given qubit-qudit state $\varrho_{AB}$ with eigenvalues $\lambda_i=\langle\psi_i|\psi_i\rangle$. Then $|\psi\rangle_{ABC}=\sum_{i=1}^{2d}|\psi_i\rangle_{AB}\otimes|i\rangle_C$ is a  purification of $\varrho_{AB}$ in a  $2\times d\times 2d$ system $ABC$. Then $\varrho_{BC}=\tr_A|\psi\rangle\langle\psi|_{ABC}$ is of rank 2 and is supported on the 2-dimensional subspace spanned by an orthornormal basis
\begin{equation}
|\phi_k\rangle_{BC}=_A\langle k|\frac1{\sqrt{\varrho_A}}|\psi\rangle_{ABC},\quad (k=0,1).
\end{equation}
Any state $\chi_{BC}$ supported on the subspace spanned by $\{|\phi_k\rangle_{BC}\}$ can be expanded $\chi_{BC}=\frac12\sum_\mu r_\mu\sigma_\mu^{BC}$  with the help of four generalized Pauli operators
\begin{equation}
\sigma_\mu^{BC}=\sum_{k,k^\prime=0}^1|\phi_k\rangle\langle k^\prime|\sigma_\mu|k\rangle\langle\phi_{k^\prime}|
\end{equation}
where coefficients $r_\mu$ ($r_0=1$) are real  $(\mu=0,1,2,3)$. For example we have
$\varrho_{BC}=(\sigma_0^{BC}+\vec x\cdot\vec \sigma^{BC})/2$. From the fact $\tr_C\sigma^{BC}_\mu=2\tr_A(\sigma_\mu\tilde\varrho_{AB})$
it follows that $\chi_B=\sum_\mu r_\mu\tr_A(\sigma_\mu\tilde\varrho_{AB})$ and
\begin{eqnarray}
&&2(1-\tr\chi_{B}^2)=2\tr[(1-V^B_{12})\chi_B\otimes\chi_B]\cr
&=&2\sum_{\mu,\nu=0}^3\tr[(1-V^B_{12})\sigma_\mu^A\otimes\sigma_\nu^A\tilde\varrho_{AB}\otimes\tilde\varrho_{AB}]r_\mu r_\nu\cr
&=&\sum_{\mu,\nu=0}^3[Q_{\tilde\varrho}]_{\mu\nu}r_\mu r_\nu,
\end{eqnarray}
in which $Q_{\tilde\varrho}$ has been defined in Eq.(\ref{Q}). The proof of the theorem is divided into the following four steps.

{\it Lemma } i) $E_F(\varrho_{BC})\ge co(1-\mC^2_{BC})$, ii) $\mC_{BC}^2=1-{\mathcal L}$, iii)
$\tau_{BC}=2(1-\tr\varrho_B^2)-(1-x^2)t_1(\tilde\varrho)$,
and iv) For a two-qubit state $\varrho_{AB}$ the bound $$D_A(\varrho_{AB}|\vec m_{\tilde\varrho})\le h(1-\tau_{BC})+S(\varrho_A)-S(\varrho_{AB})$$ holds true. The entanglement of formation, concurrence, and tangle of $\varrho_{BC}$ are defined to be, respectively,
\begin{eqnarray}
E_F(\varrho_{BC})&=&\min_{\{p_i,\pi_i^{BC}\}}\sum_ip_i S(\pi_i^B),\\
\mC_{BC}&=&\min_{\{p_i,\pi_i^{BC}\}}\sum_ip_i\sqrt{2(1-\tr(\pi_i^B)^2)},\\
\tau_{BC}&=&\min_{\{p_i,\pi_i^{BC}\}}\sum_i2p_i(1-\tr(\pi_i^B)^2),
\end{eqnarray}
where the minimization is taken over all possible ensembles $\varrho_{BC}=\sum_ip_i\pi_i^{BC}$ with $\pi_i^{BC}$ being pure and $\pi_i^B=\tr_C\pi_i^{BC}$.

{\it Proof. } i) Let $\{\hat p_i,\hat\pi_i^{BC}\}$ be the optimal ensemble for $E_F(\varrho_{BC})$ and we obtain
\begin{eqnarray}
E_F(\varrho_{BC})&=&\sum \hat p_iS(\hat\pi_i^B)\cr
&\ge&\sum \hat p_ico(2\tr(\hat\pi_i^B)^2-1) \cr
&\ge&co\left(1-\left(\sum_i\hat p_i\sqrt{2(1-\tr(\hat\pi_i^B)^2)}\right)^2\right)\cr
&\ge&co(1-\mC^2_{BC}),
\end{eqnarray}
in which the first inequality is due to $S(\varrho)\ge co(2\tr\varrho^2-1)$ the second and third inequalities are due to the fact that $co(1-z^2)$ is a convex and increasing function of $z$, respectively.

ii) Since $\varrho_{BC}$ is of rank 2, its concurrence $\mC_{BC}^2={2(1-\tr\varrho_B^2)-(1-x^2)q_2(\tilde\varrho})$ can be exactly evaluated according to {\it Theorem 4} in \cite{ulm},
where $q_2(\tilde\varrho)$ is the second largest solution to the equation $\det(Q_{\tilde\varrho}-q \eta)=0$.  Since the local filter $\sqrt[4]{1-x^2}/\sqrt{2\varrho_A}$ acting on qubit $A$ brings $Q_\varrho$ to $(1-x^2)Q_{\tilde\varrho}$ and induces an LT to $Q_\varrho$ we have $(1-x^2)q_2(\tilde\varrho)=q_2(\varrho)$ and consequently $\mC_{BC}^2=1-{\mathcal L}$.

iii) For an arbitrary state $\chi_{BC}=\frac12\sum_\mu r_\mu\sigma_\mu^{BC}$, with $\chi_B$ and $\chi_C$ denoting the reduced density matrices, we introduce a $4\times 4$ matrix $M$ whose elements are given by
\begin{eqnarray}
&&1-\tr\chi_{B}^2-\tr\chi_C^2+\tr \chi_{BC}^2\cr
&=&\tr[(1-V^B_{12})(1-V^C_{12})\chi_{BC}\otimes\chi_{BC}]\cr
&=&\frac14\sum_{\mu,\nu=0}^3\tr[(1-V^B_{12})(1-V^C_{12})\sigma^{BC}_\mu\otimes\sigma^{BC}_\nu]r_\mu r_\nu\cr
&:=&\sum_{\mu,\nu=0}^3 M_{\mu\nu}r_\mu r_\nu,
\end{eqnarray}
According to {\it Theorem 1} in \cite{osb} we have $\tau_{BC}=1-\tr\varrho_B^2-\tr\varrho_C^2+\tr\varrho_{BC}^2+(1-x^2)t$ where $t$ is the smallest eigenvalue of  the $3\times 3$ matrix $M^{3\times 3}$ obtained by deleting the first row and column of the $4\times 4$ matrix $M$ defined above. It turns out that $M=Q_{\tilde\varrho}+\frac12(1-a)\eta$ where
$$a=\tr [(V_{12}^B-V_{12}^C)|\phi_0\rangle\langle\phi_0|\otimes
|\phi_1\rangle\langle\phi_1|]=\frac{\tr\varrho_B^2-\tr\varrho_C^2}{1-\tr\varrho_{BC}^2}.$$
Thus $t=-t_1(\tilde\varrho)+(1-a)/2$ which proves Lemma iii.

iv) Let $\{p_\pm,\pi_\pm^{BC}=(\sigma_0^{BC}+\vec e_\pm\cdot\vec\sigma^{BC})/2\}$ be an optimal decomposition of $\varrho_{BC}$ for the tangle $\tau_{BC}$ and we have  $\vec e_+-\vec e_-\propto \vec e$, the eigenvector of $T_{\tilde\varrho}T_{\tilde\varrho}^T$ corresponding to its largest eigenvalue $t_1(\tilde\varrho)$, and $\vec x=p_+\vec e_++p_-\vec e_-$. If $\vec m_{\tilde\varrho}$ is the unit vector along the direction $\sqrt{1-x^2}\vec e_\perp+\vec e_\parallel$ then  we have $p_\pm=(1\pm \vec x\cdot\vec m_{\tilde\varrho})/2$ and $p_\pm\vec e_\pm=\tr(\vec\sigma \sqrt{\varrho_A}(1\pm \vec m_{\tilde\varrho}\cdot\vec\sigma)\sqrt{\varrho_A})/2$. Therefore if we measure the observable $\vec m_{\tilde\varrho}\cdot\vec\sigma$ on the qubit $A$ we obtain outcome $\pm$ with probability $p_\pm$ leaving qubit $B$ in the state $\varrho_{B|\pm}=\tr_A(1\pm \vec x\cdot\vec m_{\tilde\varrho})\varrho_{AB}/(2p_\pm)$. Thus
\begin{eqnarray}
\sum p_\pm S(\varrho_{B|\pm})&=&\sum p_\pm S(\pi^{B}_\pm)\cr
&=&\sum p_\pm h(2(\tr(\pi^{B}_\pm)^2-1)\cr
&\le& h(1-\tau_{BC})
\end{eqnarray}
in which the second equality is due to the fact that $\pi_\pm^B$ is a qubit state and the inequality is due to the fact that $h(1-z)$ is a concave function of $z$. It follows immediately that
\begin{eqnarray}
D_A(\varrho|\vec m_{\tilde\varrho})&=&\sum p_\pm S(\pi^{B}_\pm)+S(\varrho_A)-S(\varrho_{AB})\cr
&\le& h(1-\tau_{BC})+S(\varrho_A)-S(\varrho_{AB}).
\end{eqnarray}

{\it Proof of the theorem. }The upper bound is trivially true. Because of the Kaishi-Winter relation \cite{rel}
$D_A(\varrho_{AB})+S(\varrho_{AB})-S(\varrho_{A})=E_F(\varrho_{BC})$, Lemmas i and ii lead to the lower bound in the first statement. In the case of two qubits, since $\mC_{BC}^2\le 1$ leading to ${\mathcal L}\ge 0$, the lower bound of $D_A(\varrho)$ in Eq.(\ref{bds}) becomes $h({\mathcal L})+S(\varrho_A)-S(\varrho_{AB})$.  From Lemma iv
we see that $h(1-\tau_{BC})+S(\varrho_A)-S(\varrho_{AB})$ is an upper bound of $D_A(\varrho|\vec m_{\tilde\varrho})$ and therefore also an upper bound of $D_A(\varrho)$.  If $q_2(\tilde\varrho)=t_1(\tilde\varrho)$ then $\tau_{BC}=\mC^2_{BC}=1-{\mathcal L}$ because of Lemma iii, which means that the upper and lower bound coincide.

\end{document}